\documentclass[10pt,a4paper]{article}
\usepackage{jheppub_kim}
\usepackage{pdflscape}
\usepackage{amsmath}
\usepackage{amssymb}
\usepackage{dcolumn}
\usepackage{bm}
\usepackage{color}
\usepackage{amsmath}
\usepackage{breqn}
\usepackage{graphicx}
\usepackage{epsfig}
\usepackage{caption,subcaption}
\usepackage{amsfonts}
\usepackage{graphicx}
\usepackage{parskip}
\usepackage{dcolumn}
\UseRawInputEncoding

\begin{document}
\title{Warm Inflation with Barrow Holographic Dark Energy}

\author[a,b]{Moli Ghosh,}
\author[c]{Prabir Rudra,}
\author[d]{Surajit Chattopadhyay,}
\author[e,f,g,h]{Behnam Pourhassan}

\affiliation[a]{Department of Mathematics, Amity University, Kolkata, Major Arterial Road, Action Area II, Rajarhat, New Town, Kolkata 700135, India.}

\affiliation[b]{Department of Mathematics, Mrinalini Datta Mahavidyapith, Kolkata-700051, India.}

\affiliation[c] {Department of Mathematics, Asutosh College,
Kolkata-700 026, India}
\affiliation[d] {Department of Mathematics, Amity University, Kolkata, Major
Arterial Road, Action Area II, Rajarhat, New Town, Kolkata 700135,
India.}

\affiliation[e]{School of Physics, Damghan University, Damghan 3671645667, Iran.}
\affiliation[f]{Center for Theoretical Physics, Khazar University, 41 Mehseti Street, Baku, AZ1096, Azerbaijan.}
\affiliation[g]{Centre for Research Impact \& Outcome, Chitkara University Institute of Engineering and Technology, Chitkara University, Rajpura, 140401, Punjab, India.} 
\affiliation[h]{Canadian Quantum Research Center 204-3002 32 Ave Vernon, BC V1T 2L7 Canada.}

\emailAdd{molig2018@gmail.com; moli.ghosh@s.amity.edu}
\emailAdd{prudra.math@gmail.com}
\emailAdd{schattopadhyay1@kol.amity.edu}
\emailAdd{b.pourhassan@du.ac.ir}

\abstract{In this work we study the warm inflation mechanism in the presence of the Barrow holographic dark energy model. Warm inflation differs from other forms of inflation primarily in that it assumes that radiation and inflaton fields exist and interact throughout the inflationary process. After the warming process, energy moves from the inflaton to the radiation as a result of the interaction, keeping the cosmos warm. Here we have set up the warm inflationary mechanism using Barrow holographic dark energy as the driving agent. Warm inflation has been explored in a highly dissipative regime and interesting results have been obtained. It is seen that the Barrow holographic dark energy can successfully drive a warm inflationary scenario in the early universe. Finally, the model was compared with the observational data, and compliance was found.}

\maketitle

\section{Introduction}
The Big Bang model is the most successful model in modern cosmology. Despite its success, the Big Bang theory \cite{ar1,ar2,ar3} may be incomplete in its classic form because it is not sufficient to solve some cosmological problems such as the flatness problem, horizon problem, the magnetic monopole problem \cite{ar4,ar5,ar6,ar7,ar8} and the accelerated expansion of the universe \cite{sn1, sn2}. The flatness, horizon, and magnetic monopole problem can be explained by the concept of inflation which was first proposed by A. Guth in 1981 \cite{ar5}. Following inflation, the universe will enter a reheating phase, in which the inflaton decays into light particles, thermalizing the universe. By considering the cosmic expansion history from the time when the observed CMB scales escape the Hubble boundaries during inflation to when they re-enter it at a later time, it is conceivable to establish a link between the parameters of inflation and reheating. A review of inflationary cosmology can be found in \cite{review1}. Other notable developments on cosmological inflation can be found in \cite{dev1, dev2, dev3, dev4, dev5, dev6, dev7, dev8}.

The main idea of warm inflation \cite{ar9} is that unlike cold inflation (the original inflation theory), there is an extra "friction term" that acts as a regulator to fix the number density. In ordinary inflation, particle density exponentially reaches zero after inflation due to de-Sitter-type expansion. However, in warm inflation, we have specific characteristic temptation scales, such that the particle number does not go to zero due to coupling with another field.  Warm inflation differs from other forms of inflation primarily in that it assumes that radiation and inflaton exist and interact throughout the inflationary process. After the warming process, energy moves from the inflaton to the radiation as a result of the interaction, keeping the cosmos warm. Consequently, warm inflation can lead to a very smooth phase of the universe, a radiation era, offering a novel solution to the graceful exit problem. One of the biggest challenges to developing a workable inflationary model is the $\eta$ problem \cite{eta2, eta3, eta4}. Long-term slow-roll inflation is impossible due to Planck mass-suppressed adjustments to the inflaton potential, which typically result in inflaton masses of the order of the Hubble scale and yield contributions of order unity to the $\eta$ slow roll parameter. It should be noted that the cold inflation $\eta$ problem makes the flatness problem difficult to resolve. Warm inflation can naturally tackle the $\eta$ problem, hence offering much better reasoning for the flatness problem \cite{eta}. This is a great motivation for studying warm inflation.

It should also be noted that just like cold inflation affects the curvature perturbation, which in turn reciprocates to the CMB anisotropy, for warm inflation, similar calculations have been performed in \cite{ar10}. This shows that during the  warm inflation time scale (typically 16 to 60 e-fold timing), one can work with some models motivated by field theory and string theory, which can indeed give the bound on the dissipation parameter (in adiabatic approximations) \cite{ar11}. A study of warm inflation in both weak and strong dissipative regimes have been performed by Moss and Berera \cite{ar12, new1} and as a consequence, one can show how phenomenological quantities such as scalar power spectrum behaves as a function of the dissipative parameter \cite{ar13}.

To verify the warm inflation model with observation, there is a huge problem with the dissipation term, which maintains the equilibrium between the inflation field and the heat bath. The adiabatic approximation, which we use to maintain the equilibrium, would break down after the inflationary period (after 60 e-fold timing) as the characteristic mass of the radiation field is almost zero or negligible. So the scalar field perturbation would overshoot the approximation \cite{ar14}. One way to circumvent this problem is to use the heavy super potential \cite{ar15}, which can be originated via brane-antibrane stacks in string theory \cite{ar16} or extra-dimensional compactification like Kaluza-Klein theory \cite{ar17}. This bound on the heavy potential and dissipation parameters has been given via analyzing WIMP data in \cite{ar18}. Finally, we note that there is an alternative way to solve the radiation era problem by noting that we can consider the inflation field to be a pseudo-Nambu-Goldstone Boson field and during the radiation era the perturbation would not break down the approximation due to spontaneous symmetry breaking mechanism which is given in \cite{ar20,ar21}. For an overall review (with historical anecdotes) one can look into Berera's article on warm inflation \cite{revw1} and also  Rangarajan's article \cite{revw2}.

Holographic dark energy is an alternative theory of dark energy, where we attempt to apply the holographic principle to the dark energy problem. Gerard 't Hooft proposed the holographic principle \cite{ar25,ar26} inspired by the investigation of black hole thermodynamics \cite{ar24}. The relationship between a quantum field theory's greatest length and its ultraviolet cutoff \cite{ar27} can result in holographic vacuum energy, which forms dark energy on cosmological scales \cite{ar28, ar29}. The holographic principle states that all of the information contained in a volume of a space can be portrayed as a hologram, that corresponds to a theory lying on the boundary of that space. The concept of holographic principle has been widely used in various fields of physics such as in nuclear physics to study the problems of quark-gluon plasma \cite{ar30}, in the field of condensed matter to study the problems of strongly correlated systems  \cite{ar31}, in the field of theoretical physics that lead to the idea of holographic entanglement entropy\cite{ar32}, in the field of cosmology to discuss the nature of de-Sitter space and inflation \cite{ar33}. The holographic principle states that the universe's horizon entropy is proportional to its area, comparable to the Bekenstein-Hawking entropy of a black hole. This is a key step in applying it to cosmology. Applying the holographic principle to the dark energy framework, a new model of dark energy known as the Holographic dark energy model (HDE) is formed \cite{HDE1, HDE2}. In recent years Holographic dark energy has been a very lucrative candidate for explaining late-time cosmic acceleration. Rooted in the Holographic principle from string theory holographic dark energy can explain late-time cosmic acceleration and is very consistent with the observational data \cite{npb1,npb3,npb5,npb6,npb7,npb11}. The current landscape of holographic dark energy research is not just bound in general relativity, as we know that in a high-energy regime, Einstein's gravity is incapable of being consistent with quantum gravity. As a consequence, various authors have studied holographic dark energy in the context of various modified gravity such as Brans-Dicke gravity \cite{npb8}, $f(R,G)$ gravity \cite{npb10}, $f(Q)$ gravity \cite{npb9}, $f(T,T_{G})$ gravity \cite{npb12}, $f(R,T)$ gravity \cite{npb4}, $f(Q,T)$ gravity \cite{npb14}, $f(Q,C)$ gravity \cite{npb13} etc. Very recently inspired by the illustrations of the Covid-19 virus, Barrow \cite{ar34} showed that quantum-gravitational effects introduce the fractal features on the black-hole structure that leads to finite volume with infinite area. The corresponding black-hole entropy can be expressed as 
\begin{equation}\label{entropy}
S_{B}={\left(\frac{A}{A_{0}}\right)}^{1+\frac{\Delta}{2}}
\end{equation}
where $A$ is the standard horizon area , $A_{0}$ is the Planck area and $\Delta$ is the deformation parameter. It should be noted that for $\Delta=0$, the Bekenstein–Hawking entropy is recovered while $\Delta=1$ corresponds to the most intricate fractal structure. It is important to note that the aforementioned quantum-gravitationally corrected entropy differs from the standard "quantum-corrected" entropy that uses logarithmic adjustments \cite{35, 36}, however, it resembles Tsallis nonextensive entropy \cite{37, 38, 39}, however, the underlying theories and physical concepts are entirely distinct. Lastly, take note that the aforementioned effective fractal behavior is based on broad, elementary physical principles rather than particular quantum gravity computations. This increases its believability and makes it a valid initial approach to the topic \cite{ar34}. Saridakis in \cite{bhde1} used the extended Barrow relation for horizon entropy and constructed a holographic dark energy model known as the Barrow holographic dark energy (BHDE). Although BHDE possesses the usual holographic dark energy as a limit for $\Delta=0$, it is a novel scenario with a richer cosmological behavior and structure.

In \cite{warmhde} the author has studied a warm inflationary mechanism using the standard holographic dark energy and obtained very interesting results. Given the peculiar features of BHDE, we are motivated to explore a warm inflationary mechanism driven by the Barrow holographic dark energy. The fractal features inherent in BHDE are expected to produce very interesting results when incorporated in a warm inflationary mechanism. The work is organized as follows: In section 2 we discuss the warm inflationary mechanism. Section 3 is dedicated to the study of warm inflation with BHDE. Finally, the paper ends with some discussion and conclusion in section 4.

\section{Warm Inflationary mechanism}

To formulate the warm inflationary dynamics mathematically, we begin with the Einstein field equations in a spatially flat Friedmann-Robertson-Walker (FRW) universe. The metric is given by 
\begin{equation}
ds^2 = dt^2 - a^2(t)(dx^2 + dy^2 + dz^2)
\end{equation}
where $a(t)$ is the scale factor. In this framework, the total energy density comprises contributions from both the inflaton field and the radiation fluid. The evolution of this system is governed by the two fundamental equations of cosmology i.e. the Friedmann equations, 
\begin{equation}\label{c1}
H^2=\frac{1}{3 {M_{P}}^2} \left(\rho_{r}+\rho_{in}\right)
\end{equation}
\begin{equation}\label{c2}
\dot{H}=-\frac{1}{2 {M_{P}}^2}\left[\left(\rho_{r}+\rho_{in}\right)+\left(p_{r}+p_{in}\right)\right]
\end{equation}
where the subscripts '$r$' and '$in$' stands for radiation and the fluid that drives the inflation respectively. The conservation equation takes the form,
\begin{equation}\label{c3}
\dot{\rho_{in}}+3H\left(\rho_{in}+p_{in}\right)=-\Gamma \left(\rho_{in}+p_{in}\right),
\end{equation}
\begin{equation}\label{c4}
\dot{\rho_{r}}+3H\left(\rho_{r}+p_{r}\right)=\Gamma \left(\rho_{in}+p_{in}\right),
\end{equation} 
where $\Gamma$ is referred to as the dissipation coefficient, and it may be constant, dependent on the scalar field or temperature $T_{r}$, or dependent on both the scalar field and temperature.
The first slow-roll parameter is defined by,
\begin{equation}
\epsilon_{1}=-\frac{\dot{H}}{H^{2}}
\end{equation}
The next slow-roll parameters are defined by 
\begin{equation}\label{c5}
\epsilon_{n+1}= \frac{\dot{\epsilon_{n}}}{H \epsilon_{n}}
\end{equation}
There are two types of inflationary models: "warm Inflation" and "cold Inflation". In warm inflation, there is another type of slow roll parameter which is defined by 
\begin{equation}\label{c6}
\beta_{in}= \frac{\dot{\Gamma}}{H \Gamma}
\end{equation}
The evolution of the dissipation coefficient during inflationary time is represented by the parameter $ \beta_{in}$.
We define the number of e-folding $N$ between two possible values of cosmological times $t_h$ and $t_{end}$, where the time $t_h$ is the time of horizon crossing and $t_{end}$ corresponds to the end of inflation, to provide a measure of the inflationary expansion of the universe. The e-folding number in terms of the Hubble parameter can be written as
\begin{equation}\label{c7}
N=\int_{t_h}^{t_{end}} H \,dt 
\end{equation}

When there is a substantial quantity of particles during the inflationary era, warm inflation takes place. We will assume that there are sufficient particle interactions to create a thermal gas of radiation with a temperature $T$. When $T$ is greater than the energy scale determined by the expansion rate $H$ \cite{gq1, ar21,br2}, warm inflation is said to occur. The amplitude of the scalar perturbations is given by \cite{ar21,amp2}
\begin{equation}\label{c8}
\mathcal{P}_{s}=\frac{H^2}{8\pi^{2}M_{p}^{2}\epsilon_{1}}\left[1+2n_{BE}+\frac{2\sqrt{3}\pi Q}{\sqrt{3+4\pi Q}}\frac{T}{H}\right]G(Q)
\end{equation}
Here $n_{BE}$ is known as Bose-Einstein distribution which is given by $\left(e^{\frac{H}{T_{in}}}-1\right)^{-1}$. Here $T_{in}$  is the inflaton fluctuation. Also the function $G(Q)$ is represented in terms of the dissipative parameter $Q$ as \cite{gq1,warmhde}
\begin{equation}\label{c9}
G(Q)=1+0.0185 Q^{2.315}+ 0.335 Q^{1.364}
\end{equation}
It is known that both quantum and thermal fluctuations are present in case of warm inflation, and as long as $T>H$, the thermal fluctuations dominate. The function $G(Q)$ is the growth of inflation fluctuations due to the coupling to radiation and can be determined numerically \cite{ar21}. By extending the study of \cite{gq1} we numerically obtain Eq.(\ref{c9}). The two parameters, scalar spectral index ($n_{s}$) and tensor-to-scalar ratio ($r$) are widely used in inflationary scenarios. The  scalar spectral index is defined by \cite{warmhde},
\begin{equation}\label{c10}
n_{s}= \frac{d~ln(\mathcal{P}_{s})}{d ~ln (k)} +1
\end{equation}
where $k=aH$. We can write $d~ln~k=\frac{d ln k}{dN} dN$. Therefore $\frac{d ln k}{dN}$ can be expressed as dissipative parameter $Q$ and first slow roll parameter $\epsilon_{1}$ i.e $\frac{d ln k}{dN}=1- \frac{\epsilon_{1}}{1+Q}$ \cite{SPEC1,kamali}. When $Q \to 0$ and $T \to 0$  the spectral index for cold inflation reduces to $n_{s}=1-6 \epsilon+2 \eta$ (for detailed calculation see \cite{bauman}).\\

During inflationary expansion, both scalar (density) perturbations and tensor (gravitational wave) perturbations are generated. These primordial perturbations are crucial as they seed the large-scale structure formation in the universe and leave distinctive imprints on the cosmic microwave background radiation. An important observational parameter that characterizes these perturbations is the tensor-to-scalar ratio, which measures the relative strength of tensor-to-scalar perturbations. This ratio is defined as,
\begin{equation}\label{c11}
r=\frac{\mathcal{P}_{t}}{\mathcal{P}_{s}}  
\end{equation}
where $\mathcal{P}_{t}$ denotes the amplitude of the tensor perturbation which is defined by $\mathcal{P}_{t}=\frac{2 H^2}{{\pi}^2 {M_{p}}^2}$ \cite{ar21,amp2}. 

Observational evidence indicates that the scalar spectral index should lie within the range $0.96$ to $0.9684$ and the upper limit of the parameter, the tensor-to-scalar ratio is $r<0.064$ \cite{cr1}.

\section{Warm Inflation with Barrow Holographic dark energy}

In this section, we consider that inflation is driven by a holographic fluid. According to the holographic principle, holographic energy density is proportional to squared infrared cutoff $L_{IR}$. In the Barrow holographic dark energy model, the energy density is given by \cite{bhde1}.
\begin{equation}\label{d1}
\rho_{BHDE}= C L^{\Delta-2}
\end{equation}
where $C$ is the parameter whose dimension is given by $L^{-2-\Delta}$ . If $\Delta=0$ then Eq.(\ref{d1}) takes the form $\rho_{BHDE}= C L^{-2}$ that provides the standard holographic dark energy model. Here $C= 3c^2 {M_{p}}^2$ ($M_{p}$ is the Planck mass). The Granda-Oliveros (GO) cutoff is given by \cite{olivers2,olivers},
\begin{equation}\label{d2}
L^{-2}=\left(\alpha H^2+\beta \dot{H}\right),
\end{equation}
where $\alpha$ and $\beta$ are parameters. Implementing Eq.(\ref{d2}) in Eq.(\ref{d1}) we get energy density to be of the form \cite{olivers},
\begin{equation}\label{d3}
\rho_{BHDE}= 3 {M_{p}}^2 \left(\alpha H^2+\beta \dot{H}\right)^{1-\frac{\Delta}{2}}.
\end{equation}
Here $C$ is replaced by $3 {M_{p}}^2$ (considering c=1) \cite{olivers}.
 Since we have considered that inflation is driven by holographic fluid, therefore in our work $\rho_{in}=\rho_{de}$. Therefore using Eq.(\ref{d3}) in  Eq.(\ref{c1}) we get
\begin{equation}\label{d4}
H^2=\frac{1}{3 {M_{P}}^2} \left(\rho_{r}+ 3 {M_{p}}^2 (\alpha H^2+\beta \dot{H})^{1-\frac{\Delta}{2}}\right).
\end{equation}
Imposing quasi-stable production of radiation i.e $\dot{\rho}_{r}<<H \rho_{r}$ \cite{dr2,dr3} and using Eq.(\ref{c4}) and conservation equation we get
\begin{equation}\label{d5}
4H \rho_{r}= \Gamma\left(\rho_{de}+p_{de}\right).
\end{equation}
Now from Eq.(\ref{c2}) and Eq.(\ref{d5}) we arrive at
\begin{equation}\label{d6}
\rho_{r}=-\frac{3}{2}M_{p}^{2}\frac{Q}{1+Q}\dot{H}.
\end{equation}
\vspace{2mm}
The quantity $Q$ is termed as the dissipative parameter which is defined by $Q=\frac{\Gamma}{3H}$, where $\Gamma$ is the dissipation coefficient. There are two kinds of scenarios depending on the nature of $Q$:

\textbf{(I)} When $Q<1$ the standard slow-roll equation of motion of the inflaton is recovered, indicating that dissipation is not strong enough to influence the inflaton's evolution. However, the primordial spectrum of perturbations is still affected by the thermal fluctuations of the radiation energy density, which modify the field fluctuations. This is known as weak dissipative warm inflation.

\textbf{(II)} When $Q>1$, dissipation dominates both the background dynamics and the fluctuations. Because of the additional friction created by $\Gamma$, field potentials that are not flat enough to permit the typical slow-roll inflaton evolution may experience an inflationary phase. This is called strong dissipative warm inflation.

Inserting Eq.(\ref{d6}) in Eq.(\ref{d4}) $\dot{H}$ is obtained as,
\begin{equation}\label{d7}
\dot{H}= 2 \frac{1+Q}{Q} \left[\left(\alpha H^2+\beta \dot{H}\right)^{1-\frac{\Delta}{2}}-H^2\right].
\end{equation}

 From Eq.(\ref{d4}), energy density for radiation is rewritten as, 
 \begin{equation}\label{d8}
\rho_{r}=3 {M_p}^2 \left[H^2- \left(\alpha H^2+\beta \dot{H}\right)^{1-\frac{\Delta}{2}}\right].
 \end{equation}
 In particle physics models of inflation, the inflaton interacts with other fields rather than being isolated. Interactions may cause inflaton energy to dissipate into other degrees of freedom, resulting in a small percentage of the vacuum energy being converted to other forms of energy. Warm inflation involves a two-stage technique where dissipation produces particles with light degrees of freedom. In an expanding universe when relativistic particles thermalize fast enough, we can model their contribution as that of radiation,
\begin{equation}\label{d9}
\rho_{r}= \sigma_{r} T^4,
\end{equation}
 where $\sigma_{r}$ is  the Stephen-Boltzmann constant \cite{ar18} and $T$ is the temperature of the radiation field. The Stephen-Boltzmann constant can be expressed as $\sigma_{r}=\frac{{\pi}^2 g}{30}$, where $g$ is the number of degrees of freedom of the radiation field. 

 Now the temperature can be obtained by equating Eq.(\ref{d8}) and Eq.(\ref{d9}), 
\begin{equation}\label{d10}
T^4= \frac{3 M_{p}^2}{\sigma_{r}} \left[H^2-\left(\alpha H^2+\beta \dot{H}\right)^{1-\frac{\Delta}{2}}\right].
\end{equation}
We know that the first slow roll parameter is defined as $\epsilon_{1}=-\frac{\dot{H}}{H^2}$. Therefore using Eq.(\ref{d7}) the first slow roll parameter can be expressed as, 
\begin{equation}\label{d11}
\epsilon_{1}= - 2 \frac{1+Q}{Q} \left[H^{-\Delta}\left(\alpha+\beta \frac{\dot{H}}{H^2}\right)^{1-\frac{\Delta}{2 }}-1\right].
\end{equation}
From Eq.(\ref{c5}) the second slow roll parameter is obtained as,
\begin{equation}\label{d12}
\epsilon_{2}=\frac{\frac{(1+Q)}{Q}\left[(-\Delta)H^{-(\Delta+1)}\left(\alpha+\beta\frac{\dot{H}}{H^2}\right)^{1-\frac{\Delta}{2}}+H^{-\Delta}\left(1-\frac{\Delta}{2}\right)\left(\alpha+\beta\frac{\dot{H}}{H^2}\right)^{\frac{-\Delta}{2}}\left(\dot{\alpha}+\dot{\beta}\frac{\dot{H}}{H^2}+\beta\left(\frac{H\ddot{H}-2\dot{H}^2}{H^3}\right)\right)\right]-\frac{2\dot{Q}}{Q^2}}{\frac{1+Q}{Q} \left[H^{-\Delta}\left(\alpha+\beta \frac{\dot{H}}{H^2}\right)^{1-\frac{\Delta}{2 }}-1\right]H}.
\end{equation}

Dissipative effects are important during the evolution of warm inflation. Friction causes the scalar field to dissipate into a thermal bath, resulting in dissipative effects. The dissipative coefficient $\Gamma$, has been calculated from the first principles in \cite{ w1, ar15} in the context of supersymmetry. However, there are other ways to construct warm inflation which does not require the use of supersymmetry as done in \cite{kamali}. The co-efficient could be taken as constant but in a broader sense, it can be considered as a function of temperature $T$. Then, the power law form of the temperature can be considered as,
\begin{equation}\label{d13}
\Gamma= B_{t} T^{m},
\end{equation}
where $B_{t}$ is a constant. Therefore, using Eq.(\ref{d10}) $\Gamma$ can be reconstructed as,
\begin{equation}\label{d14}
\Gamma = B_{t} \left[\frac{3 {M_{p}}^2}{\sigma_{r}} \left(H^2-(\alpha H^2+\beta \dot{H})^{1-\frac{\Delta}{2}}\right)\right]^{\frac{m}{4}}.
 \end{equation}
Now substituting this result in Eq.(\ref{c6}) the slow roll parameter $\beta$ can be obtained.

\subsection{Warm Inflation in high dissipative regime}

In this subsection, we assume that inflation occurs in a high dissipative regime i.e. $Q>>1$. Now we impose the condition on Eq. (\ref{d11}) and get,
\begin{equation}\label{e1}
\epsilon_{1}=-2\left[H^{-\Delta}\left(\alpha+\beta \frac{\dot{H}}{H^2}\right)^{1-\frac{\Delta}{2}}-1\right].
\end{equation}
Using the definition of the second slow roll parameter we obtain
\begin{equation}\label{e2}
\epsilon_{2}=\frac{\left[(-\Delta)H^{-(\Delta+1)}\left(\alpha+\beta\frac{\dot{H}}{H^2}\right)^{1-\frac{\Delta}{2}}+H^{-\Delta}\left(1-\frac{\Delta}{2}\right)\left(\alpha+\beta\frac{\dot{H}}{H^2}\right)^{\frac{-\Delta}{2}}\left(\dot{\alpha}+\dot{\beta}\frac{\dot{H}}{H^2}+\beta\left(\frac{H\ddot{H}-2\dot{H}^2}{H^3}\right)\right)\right]}{\left[H^{-\Delta}\left\{\left(\alpha+\beta \frac{\dot{H}}{H^2}\right)^{1-\frac{\Delta}{2}}-1\right\}\right]H}.
\end{equation}

In our work, we assume that the parameters $\alpha$ and $\beta$ are to be of the form $\alpha=\alpha_{0} H^{\gamma}$ and $\beta=\beta_{0} H^{\delta}$, where $\alpha_0, \beta_0, \gamma, \delta$ are constants.
Imposing the condition $Q>>1$ and assuming $\alpha= \alpha_{0} H^{\gamma}$ and $\beta=\beta_{0} H^{\delta}$ Eq.(\ref{d7}) reads as,
\begin{equation}\label{e4}
\left(\alpha_{0}H^{\gamma+2}+\beta_{0} H^{\delta} \dot{H}\right)^{1-\frac{\Delta}{2}}-H^2=\frac{1}{2}\dot{H}.
\end{equation}
Truncating the higher power of $H$  from Taylor series expansion of  $(\alpha_{0} H^{\gamma+2}+\beta_{0} H^{\delta} \dot{H})^{1-\frac{\Delta}{2}}$ and using the relation in Eq.(\ref{c7}) we obtain $H$ as a function of the e-folding number $N$. Therefore we get,
 \begin{equation}\label{e5}
     H^{\delta-\gamma}={\alpha_{0}}^{1-\frac{\Delta}{2}} + C_1 e^{\frac{-2(1-p) N}{A}},
 \end{equation}
  where $A=1-2 (1-\frac{\Delta}{2}) \beta_{0} {\alpha_{0}}^{-\frac{\Delta}{2}}$ and $p=\gamma+\delta+1$. For simplicity we assume that $\delta=\frac{\Delta \gamma}{2}+\Delta$. 
  Now the slow roll parameters $\epsilon_{1}$, $\epsilon_{2}$, and $\beta_{in}$
can be obtained in terms of the e-folding number $N$.  The reconstructed slow-roll parameters in terms of $N$ is given by 
\begin{equation}\label{epp1}
    \epsilon_{1}=-2\left(-1+P^{\Delta} \left(\alpha_{0} P^{-\gamma}-K\right)^{1-\frac{\Delta}{2}}\right).
\end{equation}

The second slow roll parameter in terms of $N$ is given by,
\begin{equation}\label{ep2}
   \epsilon_{2}= \frac{\gamma_{\epsilon1}}{\gamma_{\epsilon2}},
\end{equation} 
where 
\begin{eqnarray}
\gamma_{\epsilon1}&=& C_{1} e^{D} {\alpha_{0}}^4 P^{\Delta} (\alpha_{0} P^{-\gamma}-K)^{-\frac{\Delta}{2}}\left({\alpha_{0}}^{1+\frac{\Delta}{2}} \beta_{0}(\Delta-2)\left(2 P^{-\delta} (\gamma-\delta)+C_{1}e^{D} (-2 \gamma+\Delta \gamma+2 \Delta) \right)\right)\nonumber\\
&+&C_{1} e^{D} {\alpha_{0}}^4 P^{\Delta} (\alpha_{0} P^{-\gamma}-K)^{-\frac{\Delta}{2}}\left(C_{1} e^{D} {\alpha_{0}}^{1+\Delta} P^{-\gamma} (-2 \gamma+\Delta \gamma+2 \Delta)\right)\nonumber\\
&+&C_{1} e^{D} {\alpha_{0}}^4 P^{\Delta} (\alpha_{0} P^{-\gamma}-K)^{-\frac{\Delta}{2}}\left({\alpha_{0}}^2 P^{-\gamma} \beta_{0}(\Delta-2) (-2 \gamma+\Delta \gamma+2 \Delta)\right)\nonumber\\
&-&C_{1} e^{D} {\alpha_{0}}^4 P^{\Delta} (\alpha_{0} P^{-\gamma}-K)^{-\frac{\Delta}{2}}\left((-2 \gamma+\Delta \gamma+2 \Delta) \beta_{0} P^{-\delta} 2 C_{1} e^{D} {\alpha_{0}}^4\right)\nonumber\\
&+&C_{1} e^{D} {\alpha_{0}}^4 P^{\Delta} (\alpha_{0} P^{-\gamma}-K)^{-\frac{\Delta}{2}} \left( {\alpha_{0}}^{2+\frac{\Delta}{2}} P^{-\gamma}(-2 \gamma+\Delta \gamma+2 \Delta)\right)
\end{eqnarray}

and 
\begin{equation}
\gamma_{\epsilon2}=\left((\alpha_{0}+C_{1} e^{D} {\alpha_{0}}^{\frac{\Delta}{2}})^{2}\left(-1+ P^{\Delta} \left(\alpha_{0} P^{-\gamma}-K\right)^{1-\frac{\Delta}{2}}\right) ({\alpha_{0}}^{\frac{\Delta}{2}}-2\beta_{0}\Delta \beta_{0})^{2}\right)
\end{equation}
\\
Finally warm inflation parameter $ \beta_{in}$ is given by
 \begin{equation}\label{ep3}
\beta_{in}=\frac{\gamma_{\beta_{1}}}{\gamma_{\beta_{2}}}
\end{equation}

where 
\begin{eqnarray}
\gamma_{\beta_{1}}=C_{1} e^{D} m ~\alpha_{0} P^{-2} \left(X \right),
\end{eqnarray}
with
\begin{eqnarray*}
X&=&-4 (\alpha_{0}+C_{1} e^{D} {\alpha_{0}}^{\frac{\Delta}{2}})({\alpha_{0}}^{\frac{\Delta}{2}}+ \beta_{0}(\Delta-2))\\
&-&\left(P^{-2}(\alpha_{0} P^{-\gamma}-K)\right)^{-\frac{\Delta}{2}} \left(Y\right) (\Delta-2),
\end{eqnarray*}
and
\begin{eqnarray*}   
Y&=&{\alpha_{0}}^{2+\frac{\Delta}{2}} P^{-\gamma}(2+\gamma)+C_{1} e^{D} {\alpha_{0}}^{1+\Delta} P^{-\gamma}(2+\gamma) - 2 C_{1} e^{D} P^{-\delta} \beta_{0} (2+\delta)\\
&+&\left(2 P^{-\delta} \beta_{0} (\delta-\gamma)+C_{1} e^{D} P^{-\gamma} (2+\gamma) (\Delta-2)\right)+ {\alpha_{0}}^{2} P^{-\gamma} \beta_{0} (2+\gamma) (\Delta-2).
\end{eqnarray*}

And 

\begin{equation}
  \gamma_{\beta_{2}}=\left[4 (\alpha_{0} +C_{1} e^{D} {\alpha_{0}}^{\frac{\Delta}{2}})^{2} \left(P^{-2}-(P^{-2}(\alpha_{0} P^{-\gamma}-K))^{1-\frac{\Delta}{2}}\right) \left({\alpha_{0}}^{\frac{\Delta}{2}}+\beta_{0} (\Delta-2)\right)^{2}\right]  
\end{equation}

Now we will deduce the scalar spectral index and the tensor-to-scalar ratio. Since we are in HDR the dissipative parameter $Q$ is bigger than one. Therefore, the function $G(Q)$ in Eq.(\ref{c8}) is approximated as $G(Q) \approx 0.0185 Q^{2.315}$ \cite{warmhde}. Using Eq.(\ref{c10}) we obtain  the scalar spectral index as \cite{warmhde},
\begin{equation} \label{c10a}
   n_{s}=1 + 1.815 \epsilon_{1}-\epsilon_{2} + 3.815 \beta.
\end{equation}
Again the tensor-to-scalar ratio  is obtained by computing Eq.(\ref{c8}) and Eq.(\ref{c11}) as \cite{warmhde},
\begin{equation}\label{eqr}
    r= 16 \epsilon_{1} \left(\sqrt{3} \pi \frac{T}{H} 0.0185 Q^{2.815}\right)^{-1}.
\end{equation}
Therefore the tensor-to-scalar ratio (from Eq.(\ref{eqr})) becomes,  
\begin{equation}\label{tsr2}
r=- \frac{\gamma_{r1}}{\gamma_{r2}},
\end{equation}
where
\begin{equation}
    \gamma_{r1}= \left[428.117 P^{-1}(-1+P^{\Delta}(\alpha_{0}P^{-\gamma}-K)^{1-\frac{\Delta}{2}})\right],
\end{equation}
and 
\begin{eqnarray}
\gamma_{r2}&=& \left(3^{-1+\frac{m}{4}} P\left(\frac{{M_P}^2}{\sigma_{r}} \left(P^{-2}- P^{-2} (\alpha_{0} P^{-\gamma}-K)\right )^{1-\frac{\Delta}{2}}\right)^{\frac{m}{4}}\right)^{2.8}\nonumber\\
&\times&\left(\frac{{M_P}^2}{\sigma_{r}} (P^{-2}- (P^{-2} (\alpha_{0} P^{-\gamma}-K))^{1-\frac{\Delta}{2}})\right)^{\frac{1}{4}}
\end{eqnarray}

Using Eq.(\ref{c10a}) the scalar spectral index is given by, 
\begin{equation}\label{tsr1}
    n_{s}=\frac{\gamma_{n_{1}}}{\gamma_{n_{2}}},
\end{equation}
where
\begin{eqnarray*}
   \gamma_{n_{1}}  = 1-3.63\left(-1+P^{\Delta}(\alpha_{0} P^{-\gamma}-K)^{1-\frac{\Delta}{2}}\right)-1.9075 C_{1} m P^{-2} \left(\frac{2{\alpha_{0}}^{\frac{\Delta}{2}}+2 \alpha_{0} \beta_{0}(\Delta-2)}{e^{-D} \alpha_{0}+C_{1} {\alpha_{0}}^{\frac{\Delta}{2}}}\right.
\end{eqnarray*}  
\begin{eqnarray*}
 + 2\left({\alpha_{0}}^{\frac{\Delta}{2}}+2 \alpha_{0} \beta_{0}(\Delta-2)\right)^{-2} (P^{-2}(\alpha_{0} P^{-\gamma}-K))^{-\frac{\Delta}{2}} \left(C_{1} e^{D} {\alpha_{0}}^{1+\Delta} P^{-\gamma} (2+\gamma) \right.
 \end{eqnarray*}

 \begin{eqnarray*}
 \left.
 + 2 {\alpha_{0}}^{1+\frac{\Delta}{2}} P^{-\delta} \beta_{0} (\gamma-\delta) -2C_{1} e^{D} {\alpha_{0}}^{\Delta} P^{-\delta} \beta_{0} (2+\delta)+ \right.
  \end{eqnarray*}

  \begin{equation}
  \left.\left.
  {\alpha_{0}}^{2+\frac{\Delta}{2}} P^{-\gamma} (2+\gamma)(1+C_{1} e^{D} \beta_{0}(\Delta-2)) +{\alpha}^3 P^{-\gamma} \beta_{0}(2+\gamma) (\Delta-2)\right)(\Delta-2)\right)
\end{equation}
and 
\begin{eqnarray*}
\gamma_{n_{2}}= \left(\left(P^{-2}-\left(P^{-2}(\alpha_{0} P^{-\gamma}-K)\right)^{1-\frac{\Delta}{2}}\right) \left(1+{\alpha_{0}}^{1-\frac{\Delta}{2}} \beta_{0} (\Delta-2)\right)^{2}\right)-
\end{eqnarray*}

\begin{eqnarray*}      
      \left(C_{1} e^{-D} {\alpha_{0} P^{4}}^{4} (\alpha_{0} P^{-\gamma}-K)^{-\frac{-\Delta}{2}} (2 {\alpha_{0}}^{2+\frac{\Delta}{2}} )^{\delta} \beta_{0}(\gamma-\delta) (\Delta-2) C_{1} e^{D} {\alpha_{0}}^{1+\Delta }P^{-\gamma}\times\right. 
\end{eqnarray*}   

\begin{equation}
\left.(-2\gamma+\gamma\Delta+2\Delta)+{\alpha_{0}}^{2+\frac{\Delta}{2} } P^{-\gamma} \left(1+C_{1} e^{D} \beta_{0} (\Delta-2)\right)\right)
\end{equation}

In the above equations, we have used $P$, $D$, and $K$ whose values are
\begin{equation}
P= \left(C_{1} e^D +{\alpha_{0}}^{1-\frac{\Delta}{2}}\right)^{\frac{1}{\gamma-\delta}},
\end{equation}

\begin{equation}
K=\frac{2 C_{1} {\alpha_{0}}^{\Delta} P^{-\delta} \beta_{0}}{\left(e^D \alpha_{0}+ C_{1} {\alpha_{0}}^{\frac{\Delta}{2}} \right) \left({\alpha_{0}}^{\frac{\Delta}{2}}+\beta_{0} (\Delta-2)\right)},  
\end{equation}

\begin{equation}
D=\frac{2 N (\delta-\gamma){\alpha_{0}}^{\frac{\Delta}{2}}  }{{\alpha_{0}}^{\frac{\Delta}{2}}+\beta_{0} (\Delta-2)},
\end{equation}
  
In this high dissipative regime analysis, we have successfully derived the key parameters characterizing the warm inflationary dynamics driven by Barrow holographic dark energy. The slow-roll parameters $\epsilon_1$, $\epsilon_2$ and the warm inflation parameter $\beta_{in}$ have been expressed explicitly in terms of the e-folding number $N$. These expressions incorporate both the effects of the holographic nature of dark energy through the Barrow parameter $\Delta$ and the dissipative dynamics through the coefficient $\Gamma$. This theoretical framework sets the stage for a detailed comparison with observational constraints, particularly from cosmic microwave background measurements, which we shall explore in the following subsection.

An important result of our analysis is that the reconstructed parameters naturally encode the fractal characteristics inherited from the Barrow entropy formulation while maintaining consistency with the requirements of warm inflation. The high dissipative regime ($Q \gg 1$) ensures efficient energy transfer from the holographic dark energy to radiation, which is crucial for sustaining the warm inflationary phase. This efficient dissipation mechanism, combined with the modified holographic entropy-area scaling, provides a novel perspective on the early universe dynamics.

\subsection{Results}
\begin{figure}[hbt!]
\begin{center}
\includegraphics[height=2.5in]{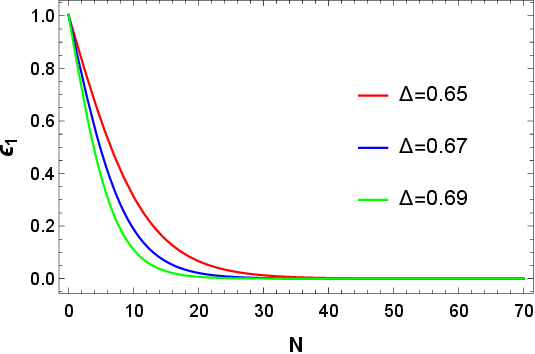}
\caption{Plot of the slow-roll parameter $\epsilon_{1}$ against the e-folding number $N$ for different values of the Barrow parameter $\Delta$. The other parameters are considered as $C_{1}=1.2$,~ $\gamma=1.02$,~$\alpha_{0}=0.06$,~ $\beta_{0}=0.01$.}
\label{figep1}
\end{center}
\end{figure}

\begin{figure}[hbt!]
\begin{center}
\includegraphics[height=2.5in]{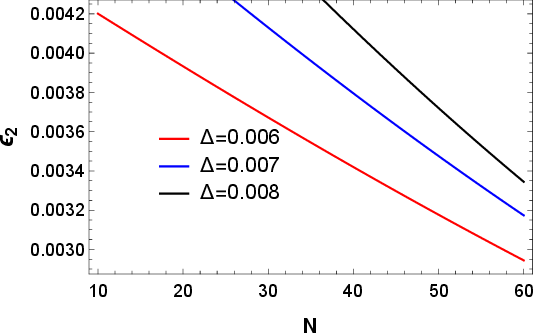}
\caption{Plot of the slow-roll parameter $\epsilon_{2}$ against the e-folding number $N$ for different values of the Barrow parameter $\Delta$. The other parameters are considered as $C_{1}=1.2$,~ $\gamma=1.02$,~$\alpha_{0}=0.06$,~ $\beta_{0}=0.01$.}
\label{EP2}
\end{center}
\end{figure}

\begin{figure}[hbt!]
\begin{center}
\includegraphics[height=2.5in]{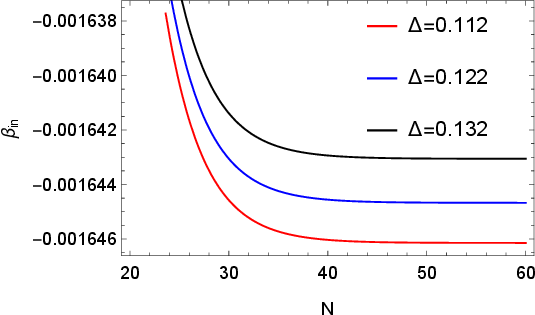}
\caption{Plot of the slow-roll parameter $\beta_{in}$ against the e-folding number $N$ for different values of the Barrow parameter $\Delta$. Other parameters are considered as $C_{1}=1.2$,~ $\gamma=1.02$,~$\alpha_{0}=0.06$,~ $\beta_{0}=0.01$.}
\label{betafig3}
\end{center}
\end{figure}

In this section, we discuss the results and compare the results with observed values to validate the model. In the previous section we have already expressed $\epsilon_{1}$, $\epsilon_{2}$, $r$, $\beta_{in}$, and $n_{s}$ in terms of e-folding number $N$. Warm inflation occurs when slow roll parameters meet the condition $\left|\epsilon_{1}\right|<<1$,~~$\left|\epsilon_{2}\right|<<1$, and~~$\left|\beta\right|<<1$. From Fig.(\ref{figep1}), Fig.(\ref{EP2}), and Fig.(\ref{betafig3}) it is apparent that all three slow roll parameters satisfy the conditions for warm inflation. The most important idea is that, if inflation stops i.e at $N=0$, the first slow-roll parameter $\epsilon_{1}$ tends to one that indicates the end of the accelerated expansion of the universe \cite{warmhde}. We have computed $\epsilon_{1}$ at $N=0$ by choosing suitable values of the parameters and we have shown that $\epsilon_{1} \approx 1$ and $\epsilon_{1}<<1$ in the region of 40-70 e-folding. Tables 1, 2, and 3 can give more insight into this where some numerical results have been presented.

\begin{figure}[hbt!]
\centering
\includegraphics[height=2.5in]{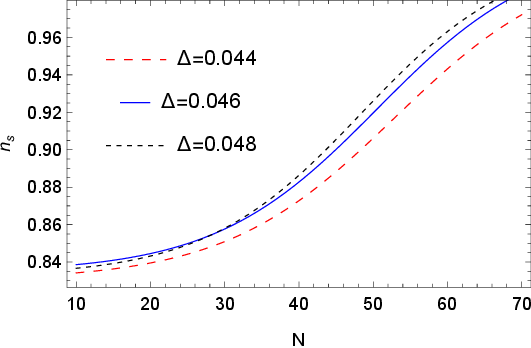}
\centering \caption{Plot of the scalar spectral index $n_s$ against the e-folding number $N$ for different values of the Barrow parameter $\Delta$. The parameters considered here are $\sigma_r=1$, $\gamma=1.02$, $\alpha_{0}=0.06$, $\beta_{0}=0.01$, $m=1$, $C_{1}=1.2$.}
\label{figns}
\end{figure}

\begin{figure}[hbt!]
\centering
\includegraphics[height=2.5in]{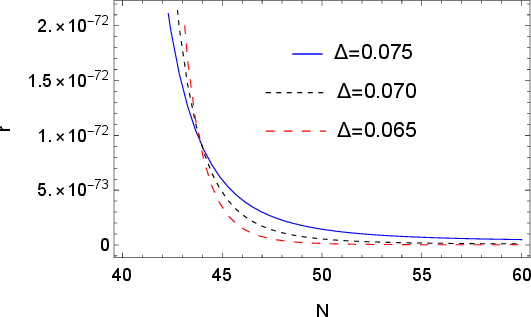}
\centering \caption{Plot of the tensor-to-scalar ratio $r$ against the e-folding number $N$ for different values of the Barrow parameter $\Delta$. The other parameters are considered as  $\sigma_r=1$, $C_1=1.2$, $\gamma=1.02$, $B_t=0.5$, $\alpha_{0}=0.06$, $\beta_{0}=0.01$, $m=1$.}
\label{figr}
\end{figure}

Fig.(\ref{figns}) illustrates spectral index $n_{s}$ versus e-folding number $N$ for different values of the Barrow parameter $\Delta$.  The latest observational data states that $n_{s}$ lies in the range $0.9642 \pm 0.0042$ \cite{cr1}. In our model, the spectral index nearly lies between the above-mentioned range. In Fig.(\ref{figr}) we have plotted the tensor-to-scalar ratio $r$ against the e-folding number $N$ for different values of the Barrow parameter $\Delta$. From the latest observational data, it is known that the upper limit of $r$ is $r<0.064$. From the plot, we see that for our model $r$ lies in this admissible range. It is also seen that generally, for a higher value of $\Delta$, we get a higher value of $r$.

\begin{table}
 \caption{The table shows numerical values of $\epsilon_{1}$ for different e-folding numbers for $\Delta=0.65$}
    \centering
    \begin{tabular}{||c|c|c|c|c|c|c|c|c||}
    \hline
        N & 0 &10  &20  &30  &40  &50  & 60 & 70\\[1ex]
        \hline
$C_{1} $ & 1.02 & 1.02 & 1.02 & 1.02 &1.02  &1.02 &1.02  & 1.02\\[1ex]
\hline
$\gamma$ &0.84  &0.84  &0.84  &0.84  &0.84  &0.84  &0.84 &0.84 \\[1ex]
\hline
$\alpha_{0}$ &1.06  & 1.06 & 1.06 & 1.06 &1.06  & 1.06& 1.06 &1.06 \\[1ex]
\hline
$\beta_{0}$ &0.01  & 0.01 & 0.01 & 0.01 & 0.01 &0.01 & 0.01 &0.01 \\[1ex]
\hline
$\epsilon_1$ &1.0035  &    0.3126 &0.066  &0.012&0.0023 &  0.00044 &0.000082 & 0.000015\\[1ex]
\hline
    \end{tabular}
   
    \label{tab:my_label}
\end{table}

\begin{table}
{\caption{The table shows numerical values of $\epsilon_{1}$ for different e-folding numbers for $\Delta=0.67$.}}
    \centering
    \begin{tabular}{||c|c|c|c|c|c|c|c|c||}
    \hline
        N & 0 &10  &20  &30  &40  &50  & 60 & 70\\[1ex]
        \hline
$C_{1} $ & 1.02 & 1.02 & 1.02 & 1.02 &1.02  &1.02 &1.02  & 1.02\\[1ex]
\hline
$\gamma$ &0.84  &0.84  &0.84  &0.84  &0.84  &0.84  &0.84 &0.84 \\[1ex]
\hline
$\alpha_{0}$ &1.06  & 1.06 & 1.06 & 1.06 &1.06  & 1.06& 1.06 &1.06 \\[1ex]
\hline
$\beta_{0}$ &0.01  & 0.01 & 0.01 & 0.01 & 0.01 &0.01 & 0.01 &0.01 \\[1ex]
\hline
$\epsilon_1$ &1.0036  &  0.1886 &0.02153  & 0.00227 & 0.00023& 0.000024  &2.60$\times 10^{-6}$ &2.72$\times 10^{-6}$ \\[1ex]
\hline
    \end{tabular}
    
    \label{tab:my_label}
\end{table}

\begin{table}
{\caption{The table shows numerical values of $\epsilon_{1}$ for different e-folding numbers for $\Delta=0.69$.}} 
    \label{tab:my_label}
    \centering
    \begin{tabular}{||c|c|c|c|c|c|c|c|c||}
    \hline
        N & 0 &10  &20  &30  &40  &50  & 60 & 70\\ [1ex]
        \hline
$C_{1} $ & 1.02 & 1.02 & 1.02 & 1.02 &1.02  &1.02 &1.02  & 1.02\\[1ex]
\hline
$\gamma$ &0.84  &0.84  &0.84  &0.84  &0.84  &0.84  &0.84 &0.84 \\[1ex]
\hline
$\alpha_{0}$ &1.06  & 1.06 & 1.06 & 1.06 &1.06  & 1.06& 1.06 &1.06 \\[1ex]
\hline
$\beta_{0}$ &0.01  & 0.01 & 0.01 & 0.01 & 0.01 &0.01 & 0.01 &0.01 \\[1ex]
\hline
$\epsilon_1$ &1.0037 &   0.11070 & 0.0068  & 0.00040 &  0.000023& 1.40$\times 10^{-6}$  & 8.28$\times 10^{-8} $ & 4.87$\times 10^{-9}$\\[1ex]
\hline
\end{tabular}   
\end{table}

\begin{figure}[hbt!]
\centering

\includegraphics[height=3.5in, width=3.5in]{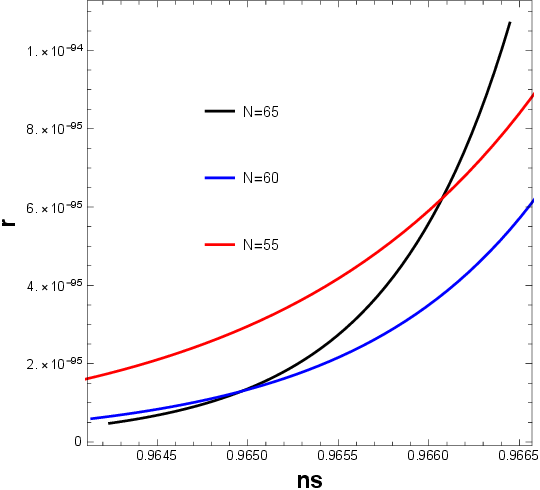}
\centering \caption{Plot of the tensor-to-scalar ratio $r$  versus the scalar spectral index $n_s$ for different values of e-folding number $N$. The other parameters are considered as $M_p=1$, $\sigma_r=1$, $C_1=1.2$, $\gamma=1.02$, $B_t=0.5$, $\beta_{0}=0.01$, $m=1$,~$\Delta=0.048$.}
\label{fignsr}
\end{figure}

\begin{figure}[hbt!]
\begin{center}
\includegraphics[height=2.5in]{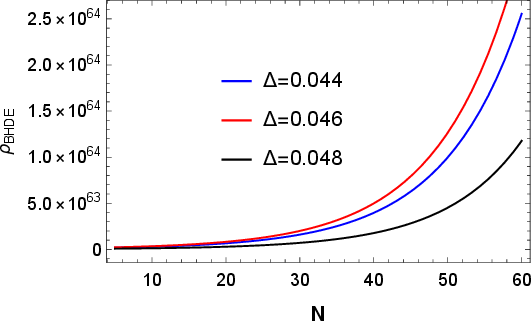}
\caption{Plot of the holographic energy density $\rho_{BHDE}$ against the e-folding number $N$ for different values of the Barrow parameter $\Delta$. The other parameters are $\sigma_{r}=1$, $C_{1}=1.2$, $\gamma=1.02$, $\alpha_{0}=0.06$, $\beta_{0}=0.01$.}
\label{rhode}
\end{center}
\end{figure}

\begin{figure}[hbt!]
\centering
\includegraphics[height=2.5in]{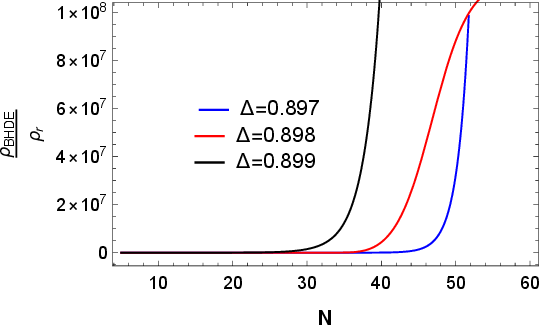}
\centering \caption{Plot of $\frac{\rho_{BHDE}}{\rho_{r}}$ against the e-folding number $N$ for different values of the Barrow parameter $\Delta$. The other parameters are considered as $M_p=1$, $\sigma_r=1$, $C_1=1.2$, $\gamma=1.02$, , $\alpha_{0}=0.06$, $\beta_{0}=0.01$.}
\label{figrhoratio}
\end{figure}

Fig.(\ref{fignsr})  illustrates tensor-to-scalar ratio $r$  versus the scalar spectral index $n_s$ for different values of e-folding number $N$. It is seen that with increasing $n_{s}$ there is a corresponding increase in $r$. Here the values of $n_{s}$ and $r$ correspond to the values at the time of horizon crossing. In Table 4 some numerical results are presented to get a better idea on these parameters. Fig.(\ref{rhode}) illustrates the behavior of energy density of BHDE for different values of the Barrow parameter. At first, it has a high value, but it decreases with the evolution of time. Inflation has an energy density of roughly $10^{64}$ GeV. Some energy was transmitted from BHDE to radiation during the inflation. We do not get a clear comparison depending on the values of $\Delta$, since there is a cross-over between the trajectories. In Fig.(\ref{figrhoratio}) the ratio of the holographic energy density and radiation density is plotted against $N$. At the initial time, the ratio was high, however, at the end of the inflation the ratio decreased and the different trajectories came close to each other. The value of the ratio is much greater than $1$, at the initial time, which shows that the inflation was driven by BHDE and $\rho_{BHDE}$ dominates. As we reach the end of the inflation we see that the ratio gets smaller, which means that the two densities approach each other.

Finally, we have observed the behavior of $T$ and $H$ in Fig.(9) and dissipative parameter $Q$ in Fig.(\ref{figQ}). In Fig.(9)  it is seen that $\frac{T}{H}$ is very much greater than 1 i.e in the presence of a thermal bath when $T > H$ the quantum fluctuations of the fields are dominated by the thermal fluctuations. So our model perfectly supports a warm inflationary scenario. In Tables 5 and 6 some numerical values of $Q$ and $T/H$ have been presented to support the idea obtained from the plots.

\begin{figure}[hbt!]
\centering
\includegraphics[height=2.5in]{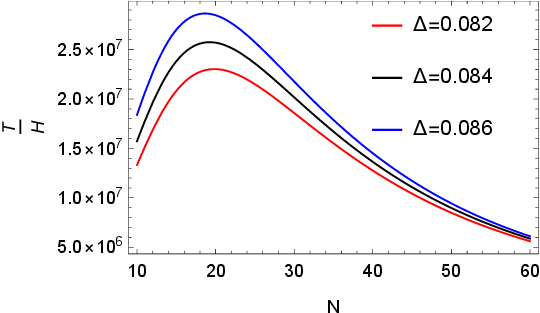}
\centering {\caption{The figure illustrate the behaviour of $\frac{T}{H}$ with respect to $N$. The parameters chosen here are  $\sigma_r=1$, $C_{1}=1.2$, $\gamma=1.02$, $\alpha_{0}=0.06$, $\beta_{0}=0.01$}.}
\label{figT}
\end{figure}

\begin{figure}[hbt!]
\centering
\includegraphics[height=2.5in]{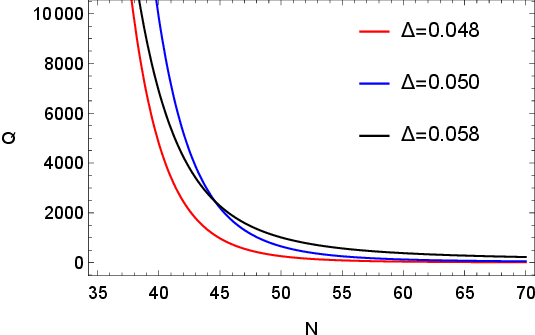}
\centering \caption{{The figure illustrate the behaviour of $Q$ with respect to $N$. The parameters chosen here are 
 $\sigma_r=1$, $C_{1}=1.2$, $\gamma=1.02$, $\alpha_{0}=0.06$, $\beta_{0}=0.01$, $m=1$}.}
\label{figQ}
\end{figure}

In Fig.(\ref{figQ}) we have plotted dissipative parameter $Q$ with respect to $N$ and we can conclude that $Q>1$ is perfectly satisfied. This indicates a strong dissipative regime.

\begin{table}[h!]
\caption{The table shows numerical values of $n_{s}$, $r$ for different values of the parameters.}
\centering
 \begin{tabular}{||c | c | c| c| c| c|c |c |c ||} 
 \hline
~~$N$~~ & ~~$\Delta$~~ & ~~$C_{1}$~~& ~~$\gamma$~~&~~$c_{t}$~~&~~$\alpha_{0}$~~&~~$\beta_{0}$~~& ~~$n_{s}$~~& ~~$r$
\\ [1ex] 
 \hline\hline
 65 & 0.044 &2.2 &0.0004 &0.5 &0.02 &0.002 &0.9591 &$2.115\times10^{-99}$  \\[1ex]
 \hline
65 & 0.046 &2.2 &0.0004 &0.5 &0.02 &0.002 &0.9675 &$1.667\times10^{-97}$   \\[1ex]
 \hline
 65 & 0.048 &2.2 &0.0004 &0.5 &0.02 &0.002 &0.9747 &$1.441\times10^{-95}$   \\[1ex]
 \hline
 65 & 0.065 &2.2 &0.0004 &0.5 &0.02 &0.002 &1.002 &$1.162\times10^{-82}$   \\[1ex]
 \hline
 65 & 0.070 &2.2 &0.0004 &0.5 &0.02 &0.002 &1.004 &$3.289\times10^{-79}$  \\[1ex]
 \hline
 65 & 0.075 &2.2 &0.0004 &0.5 &0.02 &0.002 &1.006 &$3.563\times10^{-76}$   \\[1ex]
 \hline\hline
 
60 & 0.048 &2.1 &0.0005 &0.5 &0.02 &0.002 &0.9445 &$2.702\times10^{-97}$ \\[1ex]
 \hline
60 & 0.044 &2.1 &0.0005 &0.5 &0.02 &0.002 &0.9538 &$9.232\times10^{-96}$  \\[1ex]
 \hline
 60 & 0.046 &2.1 &0.0005 &0.5 &0.02 &0.002 &0.9623 &$3.814\times10^{-97}$  \\[1ex]
 \hline
 60 & 0.065 &2.1 &0.0005 &0.5 &0.02 &0.002 &0.9992 &$1.658\times10^{-80}$  \\[1ex]
 \hline
 60 & 0.070 &2.1 &0.0005 &0.5 &0.02 &0.002 &1.0027 &$3.675\times10^{-77}$  \\[1ex]
 \hline
 60 & 0.075 &2.1 &0.0005 &0.5 &0.02 &0.002 &1.0049 &$3.518\times10^{-74}$ \\[1ex]
 \hline\hline
 
 \hline
 55 & 0.044 &1.2 &0.0002 &2.5 &0.02 &0.002 &0.9505 &$9.349\times10^{-100}$  \\[1ex]
 \hline
   55 & 0.044 &1.2 &0.0002 &2.5 &0.02 &0.002 &0.9586&$5.638\times10^{-98}$   \\[1ex]
 \hline
 55 & 0.044 &1.2 &0.0002 &2.5 &0.02 &0.002 &0.9659 &$3.541\times10^{-96}$ \\[1ex]
 \hline
  55 & 0.065 &1.2 &0.0002&2.5 &0.02 &0.002 &0.9989 &$3.203\times10^{-82}$ \\[1ex]
  \hline
   55 & 0.070 &1.2 &0.0002&2.5 &0.02 &0.002 &1.0023 &$6.647\times10^{-79}$ \\[1ex]
   \hline
    55 & 0.075 &1.2 &0.0002&2.5 &0.02 &0.002 &1.0045 &$5.945\times10^{-76}$ \\[1ex]
 
 \hline\hline
 
50& 0.044 &0.9 &0.0001 &2.5 &0.02 &0.002 &0.9452 &$9.367\times10^{-99}$ \\[1ex]
 \hline
50& 0.046 &0.9 &0.0001 &2.5 &0.02 &0.002 &0.953 &$4.892\times10^{-97}$  \\[1ex]
\hline
50& 0.048 &0.9 &0.0001 &2.5 &0.02 &0.002 &0.9602 &$2.584\times10^{-95}$  \\ [1ex]
\hline
50& 0.065 &0.9 &0.0001 &2.5 &0.02 &0.002 &0.9960 &$6.235\times10^{-82}$  \\ [1ex]
\hline
50& 0.070 &0.9 &0.0001 &2.5 &0.02 &0.002 &1.0003 &$1.061\times10^{-78}$ \\ [1ex]
\hline
50& 0.075 &0.9 &0.0001 &2.5 &0.02 &0.002 &1.0003 &$8.316\times10^{-76}$  \\ [1ex]

 \hline\hline
 
45& 0.044 &0.9 &0.00009 &4.5 &0.02 &0.002 &0.9587 &$1.621\times10^{-101}$  \\[1ex]
\hline

45& 0.046 &0.9 &0.00009 &4.5 &0.02 &0.002 &0.99648 &$2.331\times10^{-99}$  \\[1ex]
\hline

45& 0.048 &0.9 &0.00009 &4.5 &0.02 &0.002 &0.9703&$2.797\times10^{-97}$  \\[1ex]
\hline
45& 0.065 &0.9 &0.00009 &4.5 &0.02 &0.002 &0.9974&$9.544\times10^{-83}$  \\[1ex]
\hline
45& 0.070 &0.9 &0.00009 &4.5 &0.02 &0.002 &1.001&$1.876\times10^{-79}$  \\[1ex]
\hline
45& 0.075 &0.9 &0.00009 &4.5 &0.02 &0.002 &1.003&$1.571\times10^{-76}$  \\[1ex]
\hline\hline
 
 \end{tabular}
 \label{T3}
\end{table}

\begin{table}[h!]
\caption{The table shows numerical values of $Q$ for different values of the parameters.}
\centering
 \begin{tabular}{||c | c | c| c| c| c|c |c |c||} 
 \hline
~~$N$~~ & ~~$\Delta$~~ & ~~$C_{1}$~~& ~~$\gamma$~~&~~$m$~~&~~$\alpha_{0}$~~&~~$\beta_{0}$~~ & ~~$Q$\\ [1ex] 
 \hline\hline
 65 & 0.048 &2.2 &0.0004 &1 &0.02 &0.002 & 22.2046\\[1ex]
 \hline
 65 & 0.050 &2.2 &0.0004 &1 &0.02 &0.002 & 37.935\\[1ex]
 \hline
 65 & 0.058 &2.2 &0.0004 &1 &0.02 &0.002 & 282.574\\[1ex]
 \hline\hline

 60 & 0.048 &2.2 &0.0004 &1 &0.02 &0.002 & 40.4028\\[1ex]
 \hline
 60 & 0.050 &2.2 &0.0004 &1 &0.02 &0.002 & 63.3381\\[1ex]
 \hline
 60 & 0.058 &2.2 &0.0004 &1 &0.02 &0.002 & 380.529\\[1ex]
 \hline\hline

 55 & 0.048 &2.2 &0.0004 &1 &0.02 &0.002 & 90.3431\\[1ex]
 \hline
 55 & 0.050 &2.2 &0.0004 &1 &0.02 &0.002 &  127.607\\[1ex]
 \hline
 55 & 0.058 &2.2 &0.0004 &1 &0.02 &0.002 &  571.032\\[1ex]
 \hline\hline

  50 & 0.048 &2.2 &0.0004 &1 &0.02 &0.002 & 259.167\\[1ex]
 \hline
 50 & 0.050 &2.2 &0.0004 &1 &0.02 &0.002 &  326.624\\[1ex]
 \hline
 50 & 0.058 &2.2 &0.0004 &1 &0.02 &0.002 &  1012.89\\[1ex]
 \hline\hline

 45 & 0.048 &2.2 &0.0004 &1 &0.02 &0.002 & 976.335\\[1ex]
 \hline
  45 & 0.050 &2.2 &0.0004 &0.5 &0.02 &0.002 & 1101.49\\[1ex]
 \hline
  45 & 0.058 &2.2 &0.0004 &0.5 &0.02 &0.002 &2277.39 \\[1ex]

\hline\hline
 
 \end{tabular}
 \label{T3}
\end{table}

\begin{table}[h!]
\caption{The table shows numerical results of  $\frac{T}{H}$ for different values of the parameters.}
\centering
 \begin{tabular}{||c | c | c| c| c| c|c |c |c||} 
 \hline
~~$N$~~ & ~~$\Delta$~~ & ~~$C_{1}$~~& ~~$\gamma$~~&~~~~$\alpha_{0}$~~&~~$\beta_{0}$~~ & ~~$\frac{T}{H}$\\ [1ex] 
 \hline\hline
 65 & 0.082 &2.2 &0.0004  &2.06 &0.009 & $4.560 \times 10^{6}$\\[1ex]
 \hline
 65 & 0.084 &2.2 &0.0004  &2.06 &0.009 & $4.746 \times 10^{6}$\\[1ex]
 \hline
 65 & 0.086 &2.2 &0.0004  &2.06 &0.009 & $4.916 \times 10^{6}$\\[1ex]
 \hline\hline
 
60 & 0.082 &2.2 &0.0004  &2.06 &0.009 & $5.610 \times 10^{6}$\\[1ex]
 \hline
 60 & 0.084 &2.2 &0.0004  &2.06 &0.009 & $5.870 \times 10^{6}$\\[1ex]
 \hline
 60 & 0.086 &2.2 &0.0004  &2.06 &0.009 & $6.111 \times 10^{6}$\\[1ex]
 \hline\hline

 55 & 0.082 &2.2 &0.0004  &2.06 &0.009 & $6.901 \times 10^{6}$\\[1ex]
 \hline
 55 & 0.084 &2.2 &0.0004  &2.06 &0.009 & $7.258 \times 10^{6}$\\[1ex]
 \hline
 55 & 0.086 &2.2 &0.0004  &2.06 &0.009 & $7.593 \times 10^{6}$\\[1ex]
 \hline\hline

 50 & 0.082 &2.2 &0.0004  &2.06 &0.009 & $8.485 \times 10^{6}$\\[1ex]
 \hline
 55 & 0.084 &2.2 &0.0004  &2.06 &0.009 & $8.97 \times 10^{6}$\\[1ex]
 \hline
 55 & 0.086 &2.2 &0.0004  &2.06 &0.009 & $9.436 \times 10^{6}$\\[1ex]
 \hline\hline

 45 & 0.082 &2.2 &0.0004  &2.06 &0.009 & $1.041 \times 10^{7}$\\[1ex]
 \hline
 45 & 0.084 &2.2 &0.0004  &2.06 &0.009 & $1.107 \times 10^{7}$\\[1ex]
 \hline
 45 & 0.086 &2.2 &0.0004  &2.06 &0.009 & $1.171 \times 10^{7}$\\[1ex]
 \hline\hline
 
 \end{tabular}
 \label{T3}
\end{table}

\section{Discussion and Conclusion}
Warm inflation offers a framework for comprehending the dynamics of the early universe in contrast to classical cold inflation. The inflationary paradigm is introduced to rectify the shortcomings of the traditional cosmological model. Inflationary scenarios can be classified into two categories: warm inflation and cold inflation. In cold inflation, the matter field does not interact with radiation and it slowly rolls to its flat potential. However, for the warm inflationary scenario, the inflaton interacts with other fields resulting in the transmission of energy from the inflaton to the radiation field during slow-roll. The inflaton completely decays into radiation when inflation comes to an end, preventing inflation from causing the universe to enter a very cold phase. As a result, the universe enters into radiation-dominated phase without the need for a separate reheating phase.

In this article, we have studied an inflationary scenario assuming that a holographic dark fluid is the source of inflation. We have chosen Barrow holographic dark energy for this purpose in a scenario where holographic dark energy interacts with radiation and energy transmits from holographic dark energy to radiation. Here we have considered the inflationary scenario in the high dissipative regime ($Q>1$). Assuming this condition we have reconstructed the Hubble parameter as a function of the e-folding number $N$. Slow roll conditions play an important role in warm inflation. A collection of slow-roll parameters determines how consistent the slow-roll approximation is. We have shown that these parameters will satisfy the warm inflationary conditions to validate our model. Moreover, after checking the tendencies of the different inflationary parameters, it was found that there is a good agreement with the observational data \cite{cr1}. 

In a warm inflationary scenario two conditions are considered: i) thermal fluctuation dominates over quantum fluctuation i.e $T>H$, ii) Holographic density dominates radiation density. We have verified these two conditions comprehensively. We have verified that for our model thermal fluctuation dominates the quantum fluctuation. Moreover, it was also confirmed that Barrow holographic energy density was high at the time of inflation, and with the evolution of the universe, the density decreased because energy was transmitted from holographic dark energy to radiation. Finally, it was seen that $\frac{\rho_{BHDE}}{\rho_{r}}>>1 $ in the inflationary era which confirms that inflation is driven by the holographic fluid. However, the two densities come closer as inflation comes to an end. This clearly shows that the Barrow holographic dark energy model can be a novel candidate for driving warm inflation.

In a concluding note we would like to do a short discussion regarding a study presented in Ref. \cite{yokoyama} and why this is important for our study. In the Ref. \cite{yokoyama} the authors discuss the possibility of a warm inflation scenario at a generalized level and show that it is very difficult and perhaps impossible to experience inflation driven by thermal effects. The paper shows that extreme fine-tuning is required between the energy released during the expansion of the universe and the number of particles created during the warm inflation to achieve that. The authors used a complete field theoretic approach and some rigorous methods of non-equilibrium quantum statistics to show that it is very difficult to realize such a scenario. In our work, we have followed a totally different avenue to study warm inflation. It should be noted here that the main stress in our work is the application of holography to realize warm inflation which was not the case in Ref. \cite{yokoyama}. It is known that holography has its roots in black hole thermodynamics. So whenever such a concept is applied to any scenario, it is expected to bring out results that will involve stringent corrections. We think that in our work this has happened. The application of holography along with the unorthodox Barrow entropy (motivated by the fractal structure of the Covid-19 virus) may have introduced significant corrections to overhaul most of the difficulties discussed in Ref. \cite{yokoyama} to realize warm inflation. This is the reason we have observationally favorable results (observationally favorable values of the inflationary parameters $\epsilon_{1}$, $r$, $n_{s}$, etc.). Moreover, there is enough support from the literature regarding the possibility of realizing a warm inflationary scenario in the presence of a holographic dark energy. In Ref. \cite{warmhde} the author has used the standard holographic dark energy to realize a warm inflationary scenario successfully. Other relevant works on warm inflation are \cite{dr3, refo1, refo2}. These works provide enough support and motivation to explore a warm inflationary mechanism driven by holographic dark energy even when we have the knowledge presented in \cite{yokoyama}.

In future works, it would be interesting to examine our warm inflation model with Barrow holographic dark energy in the context of the Swampland conjectures. These conjectures, which provide criteria for effective field theories that can be consistently embedded in quantum gravity, have posed significant challenges for many inflationary models \cite{Vafa:2005ui, Ooguri:2006in}. Warm inflation has shown promise in addressing some of these challenges \cite{S1, S2, S3, S4, S5}, particularly regarding the de Sitter conjecture, due to its distinct dynamics and reduced requirement for slow-roll. The fractal nature of Barrow entropy could provide additional interesting features when examined under Swampland constraints. Specifically, studying how the Barrow parameter $\Delta$ affects the satisfaction of various Swampland criteria, could offer new insights into the quantum gravitational consistency of warm inflationary models.

\section*{Acknowledgments}

P.R. acknowledges the Inter-University Centre for Astronomy and Astrophysics (IUCAA), Pune, India for granting visiting associateship. The authors acknowledge the hospitality and the research facilities of IUCAA during their scientific visits. The authors also thank the anonymous referee for his/her invaluable comments that helped them to improve the quality of the manuscript. 

\section*{Data Availability Statement}

No data was generated or analyzed in this study.

\section*{Conflict of Interest}

There are no conflicts of interest.

\section*{Funding Statement}

There is no funding to report for this article.



\begin{thebibliography}{99}



\bibitem{ar1} M.Gasperini, G. Veneziano:- {\it Astroparticle Physics}  {\bf 1(3)}, 317 (1993).

\bibitem{ar2} H. Kragh:- Book chapter: Big Bang Cosmology, Book: Cosmology (pp. 371-390), CRC Press {\it eBook ISBN: 9781003418047} (2023).

\bibitem{ar3} F. Mercati:- {\it JCAP} {\bf 10} 025 (2019).

\bibitem{ar4} A. D. Linde:- {\it Phys. Lett. B} {\bf 108}, 389 (1982)

\bibitem{ar5} A. H. Guth:- {\it Phys. Rev. D} {\bf 23 }, 347 (1981).

\bibitem{ar6} A. D. Linde:- {\it Phys. Lett. B } {\bf 108}, 389 (1982).

\bibitem{ar7} R. H. Brandenberger, Inflationary Cosmology: Progress and Problems, in Large Scale Structure Formation, edited by R. Mansouri and R. Brandenberger (Springer, Dordrecht), pp. 169–211 (2000)

\bibitem{ar8} S. W. Hawking, I. G. Moss:- {\it Nuclear Physics B}, {\bf 224(1)}, 180 (1983).

\bibitem{sn1} S. Perlmutter et al.:- {\it Astrophys. J.} {\bf 517}, 377 (1999).

\bibitem{sn2} A. G. Riess et al.:- {\it Astron. J.} {\bf 116}, 1009 (1998).


\bibitem{review1} J. A. Vazquez, L. E. Padilla, T. Matos:- {\it Rev. Mex. Fis. E}, {\bf 17}, 1 (2020).

\bibitem{dev1} G. Barenboim, W. H. Kinney:- {\it JCAP} {\bf 0703}, 014 (2007).

\bibitem{dev2} M. Fairbairn, M. H. G. Tytgat:-  {\it Phys. Lett.} {\bf B546}, 1 (2002).

\bibitem{dev3} N. Nazavari, A. Mohammadi, Z. Ossoulian, K. Saaidi:- {\it Phys. Rev. D}, {\bf 93} 123504 (2016).

\bibitem{dev4} S. Maity, P. Rudra:- {\it JHAP} {\bf 2}, 1 (2022).

\bibitem{dev5} R. Maartens, D. Wands, B. A. Bassett, I. P. Heard:- {\it Phys. Rev. D} {\bf 62}, 041301 (2000).

\bibitem{dev6} S. Alexander, D. Jyoti, A. Kosowsky, A. Marciano:- {\it JCAP} {\bf (05)}, 005 (2015).

\bibitem{dev7}
E.O. Kahya, B. Pourhassan, {\it Astro. Space Science} {\bf 353}, 677-682 (2014).

\bibitem{dev8}
J. Sadeghi, B. Pourhassan, A.S. Kubeka, M. Rostami,  {\it Int. J. Mod. Phys. D}
{\bf 25}, 1650077 (2016).


\bibitem{ar9} A. Berera, L. Z. Fang:- { \it Phys. Rev. Lett.}, {\bf 74(11)}, 1912 (1995).
\bibitem{eta2} D. A. Easson, R. Gregory:- {\it Phys.Rev.D} {\bf 80} 083518 (2009)

\bibitem{eta3} S. Hardeman, J. M. Oberreuter, G. A. Palma, K. Schalm,  Ted van der Aalst:- {\it JHEP} {\bf 04} 009 (2011)
\bibitem{eta4} A. Ashoorioon, U. Danielsson, M. M. Sheikh-Jabbari:- {\it Phys.Lett.B} {\bf 713} 353 (2012)
\bibitem{eta} A. Berera;- {\it Proceedings of Science (PoS) (AHEP 2003)} {\bf 069} (2003), arXiv preprint: hep-ph/0401139, (2004).


\bibitem{ar10} S. Bartrum, M. Bastero-Gil, A. Berera, R. Cerezo, R. O. Ramos , J. G. Rosa:- {\it Phys. Lett. B}, {\bf 732}, 116 (2014).

\bibitem{ar11} A. Berera:- {\it Nuclear Physics B}, {\bf 585(3)}, 666 (2000).

\bibitem{ar12} I. G. Moss:- {\it Phys. Lett. B},  {\bf 154}, 120 (1985). 

\bibitem{new1} A Berera:- {\it Contemporary Physics} {\bf 47}, 33 (2006).

\bibitem{ar13} R. O. Ramos, L. A. da Silva:- {\it JCAP},{\bf 1303}, 032 (2013).

\bibitem{ar14}  A. Berera, M. Gleiser , R. O. Ramos:- {\it Phys. Rev. D}, {\bf 58}, 123508 (1998).

\bibitem{ar15} I. G. Moss, C. Xiong:- {\it arxiv: hep-ph/0603266}.

\bibitem{ar16} M. Bastero-Gil, A. Berera, J. G. Rosa:- {\it Phys. Rev. D}, {\bf 84}, 103503 (2011).

\bibitem{ar17} T. Matsuda:- {\it Phys. Rev. D}, {\bf 87}, 026001 (2013). 
\bibitem{ar18}  M. Bastero-Gil, A. Berera:- {\it Int. J. Mod. Phys. A}, {\bf 24}, 2207 (2009).


\bibitem{ar20} H. Mishra, S. Mohanty, A. Nautiyal:- {\it Phys. Lett. B}, {\bf 710}, 245 (2012).

\bibitem{ar21}  M. Bastero-Gil, A. Berera, R. O. Ramos, J. G. Rosa:- {\it Phys. Rev. Lett.}, {\bf 117}, 151301 (2016).


\bibitem{revw1} A. Berera:- {\it Universe} {\bf 9(6)}, 272 (2023).

\bibitem{revw2} R. Rangarajan:- {\it https://arxiv.org/pdf/1801.02648}

\bibitem{ar25} W. Fischler, L. Susskind:- {\it arxiv: hep-th/9806039}

\bibitem{ar26}  G.'t Hooft:- {\it Dimensional Reduction in Quantum Gravity}, {\it arXiv:gr-qc/9310026}.

\bibitem{ar24} J. D. Bekenstein:- {\it Phys. Rev. D}, {\bf 7}, 2333 (1973).

\bibitem{ar27} A. G. Cohen, D. B. Kaplan, A. E. Nelson:- {\it Phys. Rev. Lett.} {\bf 82}, 4971 (1999).

\bibitem{ar28} M. Li:- {\it Phys. Lett. B} {\bf 603}, 1 (2004).

\bibitem{ar29} S. Wang, Y. Wang, M. Li:- {\it Phys. Rept.} {\bf 696} 1 (2017).

\bibitem{ar30} H. Liu, K. Rajagopal, U. A. Wiedemann:- {\it JHEP} {\bf 03} 066 (2007).

\bibitem{ar31} S. A. Hartnoll:- {\it Class. Quant. Grav.} {\bf 26} 224002 (2009).

\bibitem{ar32} T. Takayanagi:- {\it Class. Quant. Grav.} {\bf 29} (2012) 153001.

\bibitem{ar33}  A. Strominger:- {\it JHEP} {\bf 10} 034 (2001).

\bibitem{HDE1}
J. Sadeghi, B. Pourhassan, and Z. Abbaspour Moghaddam, {\it Int. J. Theor. Phys.} {\bf53}, 125–135 (2014).

\bibitem{HDE2}
B. Pourhassan, A. Bonilla, M. Faizal, Everton M. C. Abreu,  {\it Phys. Dark Univ.} {\bf 20}, 41 (2018) 

\bibitem{npb1} U. K. Sharma, A. Pradhan:- {\it Mod. Phys. Lett. A} {\bf 34(13)}, 1950101 (2019).


\bibitem{npb3} V. K. Bhardwaj, A. Dixit, A. Pradhan:- {\it New Astronomy} {\bf 88}, 101623 (2021) 

\bibitem{npb5} A. Pradhan, A. Dixit:- {\it New Astronomy} {\bf 89} 101636 (2021) 

\bibitem{npb6} A. Pradhan, V. K. Bhardwaj, A. Dixit, S. Krishnannair:- {\it Int. J. Mod. Phys. A. } {\bf 36(36)}, 2150256 (2021).

\bibitem{npb7} S. Basilakos, A. Lymperis, M. Petronikolou, E. N. Saridakis:- arXiv:2312.15767. 

\bibitem{npb11} G. G. Luciano, J. Giné:- {\it Phys. Dark Universe} {\bf 41} 101256 (2023). 

\bibitem{npb8} A. Pradhan, V. K. Bhardwaj, P. Garg, S. Krishnannair:- {\it Int. J. Geom. Methods Mod. Phys.} {\bf 19(07)} 2250106 (2022)

\bibitem{npb10} K. Ghaderi, S. H. Shekh, K. Karimizadeh, A. Pradhan:- {\it Ind. J. Phys.} {\bf 98}, 2205 (2024). 

\bibitem{npb9} S. Gupta, A. Dixit, A. Pradhan:- {\it Int. J. Geom. Methods Mod. Phys.} {\bf 20(02)}, 2350021 (2023)

\bibitem{npb12} S. Gupta, A. Dixit, A. Pradhan, K. Ghaderi:- {\it Phys. Scr. } {\bf 100(1)}, 015035 (2024). 

\bibitem{npb4} G. Varshney, U. K. Sharma, A. Pradhan, N. Kumar:- {\it Chin. J. Phys. } {\bf 73}, 56 (2021).

\bibitem{npb14} N. Myrzakulov, S. H. Shekh,  A. Pradhan, K. Ghaderi:- {\it Int. J. Geom. Methods Mod. Phys.} {https://doi.org/10.1142/S0219887825500793} (2025).

\bibitem{npb13} N. Myrzakulov, S. H. Shekh, A. Pradhan:- {\it Phys. Dark Universe} {\bf 47}, 101790 (2025).
 


\bibitem{ar34} J. D. Barrow:- {\it Phys. Lett. B} {\bf 808}, 135643 (2020).

\bibitem{35} R. K. Kaul, P. Majumdar:- {\it Phys. Rev. Lett.} {\bf 84}, 5255 (2000).

\bibitem{36} S. Carlip:- {\it Class. Quant. Grav.} {\bf 17}, 4175 (2000).

\bibitem{37} C. Tsallis:- {\it J. Statist. Phys.} {\bf 52}, 479 (1988).

\bibitem{38} G. Wilk, Z. Wlodarczyk:- {\it Phys. Rev. Lett.} {\bf 84}, 2770 (2000).

\bibitem{39} C. Tsallis, L. J. L. Cirto:- {\it Eur. Phys. J. C} {\bf 73}, 2487 (2013).

\bibitem{SPEC1} S. Das, R. O. Ramos:- {\it Universe} {\bf 9(2)}, 76 (2023)




\bibitem{olivers2} L.N. Granda, A. Oliveros:- {\it Phys. Lett. B } {\bf 669(5)}, pp.275-277, 2008.

\bibitem{olivers} A. Oliveros, M.A.Sabogal,  M.A.Acero:-{\it Eur. Phys. J. C}  {\bf 137(7)}, pp.1-11,2022.

\bibitem{bhde1} E. N. Saridakis:- {\it Phys. Rev. D} {\bf 102}, 123525 (2020).

\bibitem{warmhde} A. Mohammadi:- {\it Phys. Rev. D} {\bf 104}, 123538 (2021).


\bibitem{br2} A. N. Taylor, A. Berera:- {\it Phys. Rev. D}, {\bf 62(8)}, 083517 (2000).

\bibitem{gq1} M.Bastero-Gil, A. Berera, R.O.Ramos:- {\it JCAP} {\bf (07)}, 030 (2011)

\bibitem{amp2} A. Berera, J. Mabillard, M. Pieroni, R. O. Ramos:-  {\it JCAP} {\bf (07)}, 021 )(2018).


\bibitem{bauman} D. Baumann:- {\it Cosmology (Cambridge University Press)} (2022).

\bibitem{cr1} Y. Akrami, F. Arroja, M. Ashdown, et al.:- {\it Astronomy $\&$ Astrophysics}, {\bf 641}, A10. (2020).


\bibitem{dr2}  A. Berera:- {\it Phys. Rev. D} {\bf 55}, 3346 (1997).

\bibitem{dr3} L. M. Hall, I. G. Moss, A. Berera:-  {\it Phys. Rev. D} {\bf 69(8)}, 083525 (2004).



\bibitem{w1} G. Panotopoulos, N. Videla:- {\it Eur. Phys. J. C.} {\bf 75}, 525 (2015).


\bibitem{kamali} 
V.Kamali, M. Motaharfar, and R.O. Ramos:- {\it Universe} {\bf 9(3)}, 124 (2023).

\bibitem{yokoyama} J. Yokoyama, and A. Linde:- {\it Phys. Rev. D} {\bf 60}, 083509 (1999).


\bibitem{refo1} A. Berera, T. W. Kephart:-  {\it Phys. Lett. B} {\bf 456}, 135 (1999).


\bibitem{refo2} J. M. F. Maia, J. A. S. Lima:-  {\it Phys. Rev. D} {\bf 60}, 101301 (1999).

\bibitem{Vafa:2005ui}
C. Vafa, [arXiv:hep-th/0509212]
\bibitem{Ooguri:2006in}
H. Ooguri, C. Vafa, {\it Nucl. Phys. B} {\bf 766}, 21-33 (2007).

\bibitem{S1}
J. Sadeghi, B. Pourhassan, S. Noori Gashti, and S. Upadhyay,  {\it Physica Scripta} {\bf 96}, 125317 (2021). 

\bibitem{S2}
S. Noori Gashti, J. Sadeghi, B. Pourhassan, {\it Astroparticle Physics} {\bf 139}, 102703 (2022). 

\bibitem{S3}
J. Sadeghi, B. Pourhassan, S. Noori Gashti, I Sakalli, M. R. Alipour, {\it Eur. Phys. J. C} {\bf 83}, 635 (2023).

\bibitem{S4}
J. Sadeghi, B. Pourhassan, S. Noori Gashti, E. Naghd Mezerji, A. Pasqua, {\it Universe} {\bf 8} (12), 623 (2022). 

\bibitem{S5}
S. Noori Gashti, M. R. Alipour, M. Afshar, {\bf Journal of Holography Applications in Physics} {\bf4} (3), 35-50 (2024). doi: 10.22128/jhap.2024.852.1089
\end{thebibliography}
\end{document}